\documentclass[11pt, letterpaper]{article}
\pdfoutput=1

\usepackage{amsthm}
\usepackage{amsmath}
\usepackage{amssymb}
\usepackage{amsfonts}
\usepackage{latexsym}
\usepackage{booktabs}
\usepackage{multirow}
\usepackage{graphicx}
\usepackage{natbib}
\usepackage{pdfpages}
\usepackage{algorithm}
\usepackage{algpseudocode}
\usepackage{parskip}
\usepackage[left = 3cm, right = 3cm, top = 2cm, bottom = 2cm]{geometry}
\usepackage[colorlinks,citecolor=black,urlcolor=blue]{hyperref}

\def\diag{\mathop{\rm diag}}

\def\cov{\mathop{\rm cov}}
\def\pivec{\boldsymbol{\pi}} 
\def\alphavec{\boldsymbol{\alpha}}
\def\betavec{\boldsymbol{\beta}}
\def\muvec{\boldsymbol{\mu}}
\def\phivec{\boldsymbol{\phi}}
\def\pimat{\boldsymbol{\Pi}} 
\def\phimat{\boldsymbol{\Phi}} 
\def\Imat{\boldsymbol{I}}
\def\Jmat{\boldsymbol{J}}
\def\Vmat{\boldsymbol{V}}
\def\Cmat{\boldsymbol{C}}
\def\bmat{\boldsymbol{B}}
\def\Dmat{\boldsymbol{D}}
\def\Xmat{\boldsymbol{X}}

\def\Yvec{\mathbf{Y}}
\def\yvec{\mathbf{y}}
\def\xvec{\mathbf{x}}
\def\cvec{\mathbf{c}}
\def\rvec{\mathbf{r}}

\def\pvec{\mathbf{p}}
\def\Uvec{\mathbf{U}}

\def\zerovec{\mathbf{0}}

\def\multinomial{\mathop{\rm multinomial}}

\title{Analysis of composition on the original scale of measurement}

\author{David Firth\thanks{
    Department of Statistics, University of Warwick, \href{mailto:d.firth@warwick.ac.uk}{d.firth@warwick.ac.uk}.
    ORCiD: \href{https://orcid.org/0000-0003-0302-2312}{0000-0003-0302-2312}.
   }
  \and
  Fiona Sammut\thanks{
    Department of Statistics \& Operations Research, University of Malta.
    ORCiD: \href{https://orcid.org/0000-0002-4605-9185}{0000-0002-4605-9185}.
  }
}

\date{16 December 2023}

\begin{document}
 
\maketitle

\begin{abstract}
    In current applied research the most-used route to an analysis of composition is through log-ratios --- that is, contrasts among log-transformed measurements.  Here we argue instead for a more direct approach, using a statistical model for the arithmetic mean on the original scale of measurement. Central to the approach is a general variance-covariance function, derived by assuming multiplicative measurement error.  Quasi-likelihood analysis of logit models for composition is then a general alternative to the use of multivariate linear models for log-ratio transformed measurements, and it has important advantages.  These include robustness to secondary aspects of model specification, stability when there are zero-valued or near-zero measurements in the data, and more direct interpretation.  The usual efficiency property of quasi-likelihood estimation applies even when the error covariance matrix is unspecified.  We also indicate how the derived variance-covariance function can be used, instead of the variance-covariance matrix of log-ratios, with more general multivariate methods for the analysis of composition.  A specific feature is that the notion of `null correlation' --- for compositional measurements on their original scale --- emerges naturally. 
\end{abstract} 

\emph{Keywords}:
Compositional data analysis; Generalized estimating equations; Generalized linear model;  Logit; Multiplicative error; Quasi-likelihood.

\bigskip

\section{Introduction}

\subsection{Background and setup}\label{setup}

Composition vectors represent the relative sizes of identified parts in multi-part objects.
Analysis of composition has many current applications that span physical
and biological science, the social sciences and humanities. Just a few
examples are: in electoral politics, the vote shares for different
political parties; in geology, the percentages of different minerals in
rock samples; in social and health sciences, the time-use patterns of
individuals, e.g., fractions of the day that are spent sleeping,
sedentary, physically active, etc.; and in biology, the
relative prevalence of different microbes in the human gut.  The latter
two examples --- analysis of human time-use data and of the human microbiome ---
account for much of the strong recent growth in applications of what has
become known as `compositional data analysis'.

Statistical methods in this area are dominated by highly influential work from the
1980s by John Aitchison, especially \cite{aitchison1982} and \cite{aitchison1986}; for a recent review see \citet{greenacre2023}.
The essence of Aitchison's approach is to transform observed composition
vectors to a set of contrasts among logarithms of the data (or
`log-ratios'), and then work with standard tools of multivariate
statistics such as multivariate normal distributions, linear models, and 
methods based on correlation or distance matrices.
Underlying the Aitchison approach is a notion of `compositional data'.
Suppose that \(D\) positive-valued
measurements are made on each of \(N\) objects.  The \(D\)
measurements on an object, all in the same units, relate to \(D\) separate
`parts' of the object.  Write \(y_{ik}\) for the
measurement relating to the \(k\)th part of the \(i\)th object. The
observations can then be represented as
$y_{ik} = t_i p_{ik} \ (i=1,\ldots,N;\ k=1,\ldots,D)'$,
with \(t_i = \sum_k y_{ik}\), and the unit-sum vector
\(\pvec_i=(p_{i1},\ldots,p_{iD})'\) being the observed composition of object $i$.  Aitchison's methodology operates only on the \emph{log-ratios} $\{\log(p_{ik}) - \log(p_{il})\}$, or more generally on some other defined set of linear contrasts among $\{\log (p_{i1}),\ldots,\log (p_{iD})\}$.

In practice, though, the observed $\{p_{ik}\}$ are usually error-affected \emph{measurements} of the composition(s) of interest.  Errors can come from any of the usual sources, for example sampling and other study-design effects, inaccurate instruments, numeric rounding, imperfect mixing, etc.  In any given application a more flexible approach than data-transformation, then, is to tailor a statistical model to the application --- a model that accounts properly for important sources of error, as well as targeting compositions directly through parameters.

\subsection{Extensive variables and the need to model arithmetic means}\label{extensive}

The directly measured quantities $y_{ik}$ are always \emph{extensive}, in that they have the property of physical additivity. In time-use studies, for example, the aggregate time spent on two or more related activities is the sum of the times spent on each, and the  time spent on each activity in a whole day is the sum of the times spent hour by hour.  For extensive variables it is natural, even essential, to use statistical models that target arithmetic means on the original scale; see for example \citet[][ch.~2]{cox1981} or \citet[][ch.~4]{cox2011}\null. 

A broadly useful class of models takes each $y_{ik}$ to be a realization of random variable $Y_{ik}$ with mean $E(Y_{ik})=\tau_i \pi_{ik}$, or in vector notation $E(\Yvec_i)=\tau_i\pivec_i$.  The positive-valued vector parameter $\pivec_i = (\pi_{i1},\ldots,\pi_{iD})'$ has unit sum for every $i$, and represents the composition that is measured by $\yvec_i = (y_{i1},\ldots,y_{iD})'$; and the positive scalar $\tau_i$ allows each measurement vector $\yvec_i$ to have its own expected total (or size).  This is a very general formulation based only on the first moment, and as such it allows a wide range of potential error distributions.  

Included as a special case is the most standard model for categorical \emph{counts}, namely $\Yvec_i \sim \multinomial(t_i, \pivec_i)$.  This is important not only because category counts feature in many applications, but also as a prime context where precision of measurement depends on the size parameters $\tau_i$ (which here are the known multinomial totals $t_i$).  Recent work of \citet{fiksel2022} develops methods in a similar spirit to the approach proposed here, to take account of multinomial-type sampling errors in a specific application context.   
 
Even with \emph{continuous} measurements there is often potential for more complex error structures than are supported by the use of log-ratios. For example, small samples (of rock, for instance) might be affected by physical detection limits that are irrelevant with larger samples; in time-use studies the tendency of survey respondents to report times to the nearest half-hour (say) will affect small and large time-periods quite differently; etc.  

\subsection{Focus here on multiplicative errors}

While the general first-moment specification $E(\Yvec_i) = \tau_i\pivec_i$ provides substantial flexibility to account for a wide variety of different error structures, in the remainder of this paper we focus on the particular case of \emph{multiplicative} errors (which might alternatively be called \emph{relative} or \emph{proportionate} errors).  
Specifically it is assumed that --- for each part $k$ of each object $i$ ---
$Y_{ik} = \tau_i \pi_{ik} U_{ik}$,
or in vector notation 
\begin{equation}\label{multiplicative} 
    \Yvec_i = \tau_i \pimat_i \Uvec_i, 
\end{equation}
where $\pimat_i$ denotes the matrix $\diag(\pivec_i)$ and the relative errors $U_{ik}$ 
all have unit mean.  We denote the error variance-covariance matrix by $\cov(\Uvec_i) = \phimat$, the same for all $i$; this homoskedasticity can easily be relaxed where appropriate, and we comment briefly on it in the Discussion section below.  It is natural to think in terms of the stronger assumption that the relative error vectors $\{\Uvec_i\}$ are i.i.d., but that is not a necessary assumption for what follows.

If the random multipliers $\{U_{ik}\}$ are restricted to be positive, the log-ratios $\log(Y_{ik}/Y_{il})$ are available and are free of the size parameter $\tau_i$. It is easily shown that if the relative-error vectors $\{\Uvec_i\}$ are drawn from a multivariate lognormal distribution, suitably scaled to have $E(U_{ik}) = 1$ for all $(i,k)$, then the resulting parametric model is the family of logistic normal distributions \citep[][ch.~6]{aitchison1986}.  

The general multiplicative model (\ref{multiplicative}) is more widely applicable than log-ratios, though, because the relative errors $U_{ik}$ are not restricted to be positive: zeros are allowed, or even negative values (as might be appropriate, for example, in situations where the measurement mechanism involves a differencing operation).  In addition, importantly, the multiplicative model (\ref{multiplicative}) always has parameters that relate directly to arithmetic means and totals on the original scale of measurement, regardless of any other distributional details of the unit-mean error vectors $\{\Uvec_i\}$.

\section{Variance-covariance function}\label{Vsec}

The multiplicative model (\ref{multiplicative}) describes how $\Yvec_i$ for each object $i$ is an unbiased, error-affected measurement of the corresponding mean vector $\tau_i\pivec_i$.  The part of the measurement error that relates purely to composition is $\Yvec_i - T_i\pivec_i$, where $T_i = \sum_k Y_{ik}$.  We can write
$\Yvec_i - T_i\pivec_i = (\Imat - \pimat_i \Jmat) \Yvec_i$,
where the $D\times D$ matrices $\Imat$ and $\Jmat$ are respectively the identity matrix and the matrix of ones.
Since $\cov(\Yvec_i/\tau_i) = \pimat_i\phimat\pimat_i$, the variance-covariance matrix of $(\Yvec_i - T_i\pivec_i)/\tau_i$ also is free of $\tau_i$:
\begin{equation}
\begin{split}\label{Vfun}
    \cov\left(\dfrac{\Yvec_i - T_i\pivec_i}{\tau_i }\right) 
       & =  (\Imat - \pimat_i \Jmat) \pimat_i\phimat\pimat_i (\Imat - \pimat_i \Jmat)' \\
       & =  (\pimat_i - \pivec_i\pivec_i') \phimat (\pimat_i - \pivec_i\pivec_i')\\
       & = \Vmat(\pivec_i; \phimat),\ \text{say.}
\end{split}
\end{equation}
The matrix variance-covariance function $\Vmat$ is the natural extension, beyond the case $D=2$, of a scalar variance function that was suggested previously in \citet{wedderburn1974} for a generalized linear model with continuous proportions as response variable.  The case $D=2$ has essentially univariate measurements, since $\pi_{i1}=1-\pi_{i2}$ for all $i$; and the suggestion made in \citet{wedderburn1974} was to use the variance function $V(\pivec_i; \phi) = \phi (\pi_{i1}\pi_{i2})^2$ for quasi-likelihood analysis of a logit-linear model.  Wedderburn's suggested variance function is proportional to the square of the Bernoulli variance function $\pi_{i1}\pi_{i2}$, with scalar constant of proportionality $\phi$.  The more general form (\ref{Vfun}) uses the multinomial variance-covariance function $\pimat_i - \pivec_i\pivec_i'$ to extend from Bernoulli, and the error dispersion matrix $\phimat$ in place of scalar dispersion $\phi$.  Because of this connection, we will call $\Vmat(\pivec_i; \phimat)$ the `generalized Wedderburn' variance-covariance function.

The multinomial variance-covariance matrix $\pimat_i - \pivec_i\pivec_i'$ is singular, and its Moore-Penrose pseudo-inverse was derived in \citet{tanabe1992}.  In the $D\times D$ case, and with $\Cmat = \Imat - D^{-1}\Jmat$ denoting the `centering' projection matrix that sweeps out any $D$-vector's arithmetic mean, the pseudo-inverse is
$\label{MPinv}
(\pimat_i - \pivec_i\pivec_i')^+ = \Cmat \pimat_i^{-1} \Cmat$.
A useful property of this symmetric pseudo-inverse in what follows will be that
\begin{equation}\label{stabilize}
(\Cmat \pimat_i^{-1} \Cmat) \Vmat(\pivec_i; \phimat) (\Cmat \pimat_i^{-1} \Cmat) = \Cmat \phimat \Cmat,    
\end{equation}
for all composition vectors $\pivec_i$ and error covariance matrices $\phimat$.  For any $i$, then, the pseudo-inverse $\Cmat \pimat_i^{-1} \Cmat$ can be used to stabilize the variance-covariance matrix of the compositional error vector $\Yvec_i - T_i\pivec_i$, eliminating the dependence on $\pivec_i$.

The remainder of this paper outlines some applications of the variance-covariance function $\Vmat(\pivec_i; \phimat)$ and its pseudo-inverse.  The application to regression models for compositional response, via quasi-likelihood equations that are linear in un-transformed measurements, is covered in the next section.  While section 4 indicates briefly how $\Vmat(\pivec_i; \phimat)$ and the associated variance-covariance stabilizing transformation can be used in some of the standard methods of multivariate analysis, also without the need for log-ratio transformation of compositional measurements.

\section{Compositional quasi-likelihood and logit models}


This section focuses on modeling the dependence of compositions on explanatory variables, through logit-linear regression models.  It would be possible to consider link functions other than logit here, but the logit link has some key advantages: multi-category logit models are already familiar, and their interpretation here is essentially the same as in multinomial models for discrete choice; and it turns out that the specific combination of logit link with the generalized Wedderburn variance-covariance function (\ref{Vfun}) yields an exceptionally clean quasi-likelihood analysis with appealing properties. 

Regression the other way around, with compositional measurements as explanatory or predictor variables rather than as the response, is an important topic that is not considered here.  We remark only that there is useful work on this in the older literature, as well as in a great deal of current applied research.  \citet{cox1971}, for example,  gives valuable insights and remains highly relevant.

\subsection{Compositional logit model}

Suppose that each object $i$ has associated with it a vector of $p$ covariate values, and write $\xvec_i = (1, x_{i1},\ldots,x_{ip})$ with $x_{ir}$ the value of the $r$th covariate for object $i$.  For the dependence of composition vector $\pivec_i$ on $\xvec_i$, the compositional logit model is
\begin{equation}\label{logit}
    \pi_{ik} = \dfrac{\exp(\xvec_i'\betavec_k)}{\sum_k \exp(\xvec_i'\betavec_k)}\quad (i=1,\ldots,N;\ k=1,\ldots,D),
\end{equation}
where $\betavec_k$ is a separate vector of $p+1$ unknown parameters for each part $k$ of the composition. As represented here the model is over-parameterized: the parameter vectors $\{\betavec_k:\ k=1,\ldots,D\}$ are not separately identified.  A linear constraint of the form $\cvec'\bmat = \zerovec$, where $\sum_kc_k = 0$ and $\bmat$ is the $D \times (p+1)$ matrix whose rows are the $\{\betavec_k\}$, can be used to identify the parameters.   

This is the same as the well known multinomial logit model, commonly used for response variables that are discrete choices or category counts.  We use the name `compositional logit' here only to eliminate any potential confusion with models based on multinomial distributions.

The compositional logit-linear model (\ref{logit}) has, for all $k$ and $l$ in $\{1,\ldots,D\}$,
\[
    \log(\pi_{ik}/\pi_{il}) = \xvec_i'(\betavec_k - \betavec_l).
\]
This can be compared with the corresponding multivariate linear-model formulation for log-ratios in the data \citep[][sec.~7.6]{aitchison1986}, which assumes instead that
\begin{equation*}
    E[\log(p_{ik}/p_{il})] = \xvec_i'(\betavec_k - \betavec_l).
\end{equation*}
This is the familiar difference between transformed-response and generalized linear model approaches to regression analysis. The logit model (\ref{logit}) is a model for expectations on the original scale of measurement: it uses $\log[E(p_{ik})]$ in place of $E[\log(p_{ik})]$.  

The interpretation and properties of the multinomial logit transformation are already well understood.  Notable among its properties is that the logits have `independence of irrelevant alternatives' (IIA), meaning that the model for any subset of $\{\pi_{i1},\ldots,\pi_{iD}\}$ is unaffected by the existence of the remaining parts.  This might or might not be a desirable property, depending on the application context.  For situations where IIA is an unrealistic assumption, a commonly useful alternative to the single model (\ref{logit}) is a nested sequence of such logit models for an identified hierarchy of subsets of the parts; in practice, then, the IIA property is not a limitation on applicability of the general model (\ref{logit}).

\subsection{Quasi-likelihood}

The method of quasi-likelihood estimation, as the appropriate extension of (generalized) least squares to regression models where variance depends on the mean, is well established.  The method was introduced for univariate-response models by \citet{wedderburn1974}, and extended to the multivariate case by \citet{mccullagh1983}; a useful overview is given in \citet[ch.~9]{mccullagh1989}.

The general form of quasi-likelihood estimating equations, for a model with response vector $\yvec$, regression function $\muvec(\betavec)$, and variance-covariance function $\Vmat(\muvec; \phivec)$ is
\begin{equation}\label{ql}
\Dmat' \Vmat^-(\muvec; \phivec) (\yvec - \muvec) = \zerovec,
\end{equation}
\citep{mccullagh1983}, where $\Dmat$ is the matrix of partial derivatives $\partial\muvec / \partial \betavec$, and $\phivec$ denotes any other dispersion parameters that might be involved in the dependence of $\Vmat$ on $\muvec$.  The variance-covariance function $\Vmat$ may be singular; the notation $\Vmat^-$ denotes a generalized inverse.

Here we will assume independence of the measurements on different objects $i$, an assumption that can be relaxed to handle patterns of dependence between objects, for example spatial or temporal dependence.  With the independence assumption, the quasi-score contributions are additive and the general form (\ref{ql}) becomes
\[
\sum_{i=1}^N \Dmat_i' \Vmat^-(\muvec_i; \phivec) (\yvec_i - \muvec_i) = \zerovec.
\]

In the logit-linear regression (\ref{logit}), each $i$ has $\Dmat_i = (\pimat_i - \pivec_i\pivec_i') \Xmat_i$,
where the covariate vector $\xvec_i$ has been expanded to a $D \times D(p+1)$ matrix through the Kronecker product $\Xmat_i = \Imat \otimes \xvec_i'$\null.  In order to use the generalized Wedderburn variance-covariance function from (\ref{Vfun}), a value is needed for the scale parameter $\tau_i$; after the data has been observed the natural value to use for $\tau_i$ is the observed total $t_i$, so that the compositional error for object $i$ becomes more simply expressed as $\pvec_i - \pivec_i$.  For the required generalized inverse of $\Vmat$ we use the Moore-Penrose pseudo-inverse, which is
$\Vmat^+ = (\Cmat\pimat_i^{-1}\Cmat) (\Cmat \phimat \Cmat)^+ (\Cmat \pimat_i^{-1} \Cmat)$,
with $\Cmat$ defined as in section \ref{Vsec}.  
Simplifying facts are that $(\pimat_i - \pivec_i\pivec_i')  (\Cmat\pimat_i^{-1}\Cmat) = \Cmat$, and that the compositional errors $(\pvec_i - \pivec_i)$ are already centered on account of the unit-sum constraint on both $\pvec_i$ and $\pivec_i$.  The system of $D(p+1)$ quasi-likelihood equations is then:
\begin{equation}\label{qleqn}
\sum_{i=1}^N \Xmat_i' \Cmat (\Cmat \phimat \Cmat)^+ \Cmat\pimat_i^{-1}(\pvec_i - \pivec_i) = \zerovec.
\end{equation}
These are nonlinear equations, requiring iterative solution.  The standard machinery of iterative generalized least squares (GLS) can be brought to bear, and indeed the GLS calculations reveal a further, important simplification.  Each GLS step fits a multivariate linear model having separate coefficients $\betavec_k$ for each part $k$ of the compositional response, and it is known that GLS in such a model reduces simply to ordinary least squares: the weight matrix plays no part in the solution.  This is the phenomenon sometimes referred to as `seemingly unrelated regressions', and details can be found in standard texts on multivariate statistics \citep[e.g.,][sec.~6.6.3]{mardia1979}.  The upshot for the quasi-likelihood equations (\ref{qleqn}) is that the apparent dependence on $\phimat$ is illusory; and because $\Cmat$ is idempotent the system of equations becomes simply
\begin{equation}\label{qleqn2}
\sum_{i=1}^N \Xmat_i' \Cmat\pimat_i^{-1}(\pvec_i - \pivec_i) = \zerovec.
\end{equation}

The estimating equations derived here are for a multivariate generalized linear model, with vector response, and as such they are outside the scope of standard software such as \texttt{glm()} in \emph{R} \citep{R2023}. 
A computational device that mimics the familiar Poisson-loglinear approach to fitting multinomial logit models \citep{baker1994} is as follows.  Define the linear predictor function
$\eta_{ik}(\alphavec, \betavec) = \alpha_i + \xvec_i'\betavec_k\ (i=1,\ldots,N;\ k=1,\ldots,D)$,
and then solve simultaneously the equations
\begin{equation}\label{trick}
\begin{split}
    \sum_{i=1}^N \left[\dfrac{p_{ik}}{\exp(\eta_{ik})} - 1 \right] x_{ir} & = 0\quad (\text{all } r,k) \\
    \sum_{k=1}^D \exp(\eta_{ik}) & = 1\quad (\text{all } i).
\end{split}
\end{equation}
The first set of equations in (\ref{trick}) are maximum-likelihood equations for a gamma log-linear model, while the second set merely ensures that the fitted totals are all 1\null.  It can easily be shown that the same $\hat\betavec$ that solves (\ref{trick}) also solves (\ref{qleqn2}).  This route to solving the quasi-likelihood equations is implemented, through iterative use of standard univariate regression functions, in a prototype \emph{R} package at \url{https://github.com/DavidFirth/compos}.

\subsection{Properties}

\subsubsection{Asymptotic distribution}

From the general theory of quasi-likelihood estimating equations \citep{mccullagh1983}, it is known that the resulting estimators are consistent and asymptotically normal, with asymptotic variance-covariance matrix the inverse of the quasi-information $\Dmat'\Vmat^-\Dmat$\null.  For the compositional logit model studied here, with the assumed generalized Wedderburn variance-covariance function, it emerges from straightforward algebra that the quasi-information does not depend on the parameters $\betavec$, only on the error dispersion matrix $\phimat$ and the $N \times (p+1)$ design matrix $\Xmat$ whose rows are the covariate vectors $\xvec_i$.  The asymptotic variance-covariance of the estimated parameters can be written succinctly in the form
\begin{equation}\label{vcov}
    \cov(\hat\betavec) = \Cmat\phimat\Cmat \otimes (\Xmat'\Xmat)^-,
\end{equation}
where we have again allowed the possibility of a generalized inverse as will be needed if the model is represented in the symmetric, unconstrained way that has been used above.  The `centered' dispersion matrix, $\Cmat\phimat\Cmat$, is estimated straightforwardly from the standardized residuals $\Cmat\hat\pimat_i^{-1}(\pvec_i -\hat\pivec_i)$, owing to the useful property noted at (\ref{stabilize}) above.  From (\ref{vcov}), then, approximate inference can readily be made on any identifiable parameter combination.

\subsubsection{Optimality}

Again from the standard theory \citep{mccullagh1983}, it is known that the quasi-likelihood equations are optimal among linear, unbiased estimating equations. A notable feature of the model developed here is that, because the estimating equations are the same for all error-dispersion matrices $\phimat$, the simplest version (\ref{qleqn2}) which neglects $\phimat$ is optimal whatever might be the true contents of matrix $\phimat$.  This is an appealing consequence of the `seemingly unrelated regressions' aspect of the model that was mentioned above.

\subsubsection{Robustness}

The method of quasi-likelihood has a general model-robustness property, which is that the resulting estimator remains consistent even under failure of the assumed variance-covariance function $\Vmat$.   It should be noted, though, that the variance-covariance matrix in (\ref{vcov}) is `model based', in that it assumes correctness of the multiplicative-error model.  If that assumption is in doubt --- for example if the reality might be some more complex measurement mechanism --- then the use of a `sandwich' variance-covariance estimator \citep[as in][]{liang+zeger1986} can be considered in place of (\ref{vcov}).  

\subsubsection{The `orthogonal link'}

As we have seen, the pairing of compositional logit model and generalized Wedderburn variance-covariance function results in a quasi-information matrix that is free of the regression parameters.  A specific consequence of this is that orthogonality in $\Xmat'\Xmat$ allows separate inference on the corresponding parameters; this can be helpful in the analysis of carefully planned experiments, for example.

\subsubsection{Zeros}

A major problem for the use of log-ratio methods in practice is that there is instability whenever any compositional measurement $p_{ik}$ is close to zero, and complete breakdown when any value is exactly zero.  Much research effort over the years has been spent on proposed fixes for this problem, mainly the addition of small constants to the data to move troublesome observations away from zero; unfortunately, since logarithms of small numbers are large, conclusions of analysis are often very sensitive to the specific choice of any such adjustment.

The model and method developed above have no such problems.  In essence this is because the analysis targets arithmetic means, for which zero-valued observations have no special status.  This contrasts strongly with the log-ratio approach, which in effect targets geometric means instead.

\subsubsection{The Aitchison principles}

The argument for log-ratio analysis as developed in \citet{aitchison1986} is based upon a stated set of principles that should be followed in the analysis of compositional data.  Chief among those principles are `scale invariance' and `subcompositional coherence'.  \citet{scealy2014} give an insightful account.

The multiplicative model described above has the property of scale invariance, by design.  However, the Aitchison requirement of `subcompositional coherence' --- which is like the IIA property described in section 3.1, but governed by the data instead of the model --- is violated by the analysis described above.  Our view is that this is no bad thing: if the `independence of irrelevant alternatives' notion is thought appropriate for a particular application context, it is better framed as a requirement of the models and methods to be used rather than as a requirement of the data (which are usually affected by errors).  It is the `subcompositional coherence' requirement that rules out the possibility of zeros in the data, for example; the IIA property relating to models imposes no such restriction, as we have seen.

\subsection{Example}

In \citet{aitchison1986}, Arctic lake sediment data were used to illustrate the multivariate linear model for log-ratios.  The data are the recorded composition of lake-bed sediment samples into the three parts (sand, clay, silt); and also the depth of the lake at the location of each sample, which is used as a covariate (on the log scale).  The original source for the data makes clear that sand, clay and silt are naturally ordered parts, so this is a context where a sequence of \emph{nested} logit models might be appropriate.  However, \citet[sec.~7.6]{aitchison1986} ignores that aspect, and for that reason we do the same here as our intention is a like-for-like illustrative comparison of Aitchison's log-ratio linear model with the compositional logit model.  
\begin{center}
    \includegraphics[width=3in]{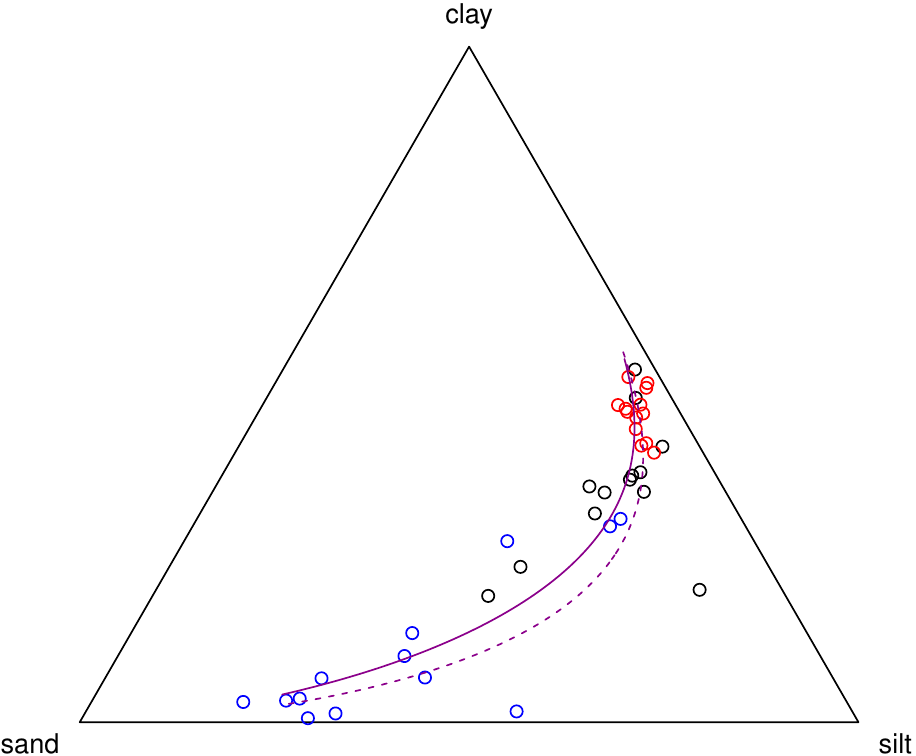}
\end{center}
The above simplex plot shows the compositional data, with blue points indicating samples taken at the shallowest depths, and red points the deepest.  The two plotted curves represent the fitted models: dashed is the log-ratio multivariate linear model given in \citet[sec.~7.6]{aitchison1986}, while the solid curve is the compositional logit model with the same single covariate log(depth).  The main difference between the two fitted curves is that the model based on log-ratios is drawn towards the points near the edge of the simplex: there are no zero-valued parts in this particular set of compositional data, but values close to zero have a strong influence on the fit of the model.

\section{Compositional covariance, correlation and distance}\label{multivariate}

In this section we indicate briefly how the approach taken in this paper can contribute to some other methods of analysis, beyond regression models.  Various standard multivariate methods are based on covariance and correlation matrices, and others are based on distance measures. We make suggestions here based on based on the above development, which we hope might stimulate further work.

\subsection{Covariance and correlation}

Variances and covariances in multivariate analysis concern departures from a mean vector.  In the context of the multiplicative model (\ref{multiplicative}), the implication is that we should consider only information on the distribution of the relative error vectors $\Uvec_i$, and in particular on the error variance-covariance matrix $\phimat$.

The key to this is equation (\ref{stabilize}) above, from which it follows that the empirical variance-covariance matrix of what might be called the `standardized residual' vectors $\{\rvec_i = \Cmat\hat\pimat^{-1}(\pvec_i - \hat\pivec):\ i=1,\ldots,N\}$ is an estimate of $\Cmat\phimat\Cmat$.  Here $\hat\pivec$ denotes the arithmetic mean, $\sum_i \pvec_i / N$, and $\hat\pimat$ the corresponding diagonal matrix.

The analogous construction in the log-ratio framework is the `centred logratio covariance matrix' \citep[defn.~4.6]{aitchison1986}, which instead uses, on the log scale, departures from the \emph{geometric} mean.

A specific point to note is that an estimate of $\Cmat\phimat\Cmat$ such as the one suggested here can be used to assess proximity to `null correlation' on the original scale of measurement.  This was identified in \citet[sec.~3.3]{aitchison1986} as a difficult problem.  When $\phimat$ is diagonal, corresponding to null correlation, the matrix $\Cmat\phimat\Cmat$ has a special structure determined by only the $D$ variances in $\phimat$; and departure from such structure can be tested, if required.

\subsection{Distances}

Use of the above covariance matrix leads naturally to (squared) Mahalanobis distances of the form
$d^2_{\phimat}(\pvec_i,\pvec_{j}) = (\rvec_{i}-\rvec_j)^{'}(\Cmat\phimat\Cmat)^{+}(\rvec_i-\rvec_j)$\null, based on the standardized residual vectors $\{\rvec_i\}$\null. Alternatively, for a distance measure that does not need an estimate of $\phimat$, we could simply use $d^2_{\Imat}(\pvec_i,\pvec_{j}) = (\rvec_{i}-\rvec_j)^{'}(\rvec_i-\rvec_j)$.  The latter is similar to the so-called `Aitchison distance' measure from \citet{aitchison1992}, which has the same form but based on values of $\log(p_{ik}/\pi_{k})$ rather than $p_{ik}/\pi_{k}-1$, and with geometric means replacing arithmetic means in the estimation of $\pivec$. Thus the new suggestion $d_{\Imat}$ will be similar to Aitchison distance when compositions are close to one another. But $d_{\Imat}$, as indeed the more general $d_{\phimat}$, presents no problem in handling \emph{zeros} in the data; whereas zeros are a major obstacle in the practical use of Aitchison distance \citep[e.g.,][]{stewart2017}. 

\section{Discussion}

A key aim of this work is to argue that analysis of composition is a standard kind of statistical activity, best understood through statistical models that account for sampling and measurement processes and that have parameters defined in such a way that composition is targeted.  The specific development of the multiplicative model above shows that, even where measured totals are deemed irrelevant, it is possible to make progress without log-ratio transformation of the data.  Nothing in the present paper is prescriptive: the intention is to open up this area of statistical methodology, which is increasingly important in applied research, to the development of flexible statistical models and methods for specific types of application.  Recent work of \citet{scealy2023} develops a distinct approach in a similar spirit.

We conclude by mentioning briefly just a few of the ways in which the development of sections 2--3 might be extended:
\begin{enumerate}
\item 
    The approach is easily generalized to handle situations where different `objects' in the data carry information on different parts of the composition of interest.  This is similar to the use of a reduced choice set in the discrete case --- an extreme example of which is pair-comparison data.
\item 
    The assumption of homogeneous dispersion $\phimat$ for the multiplicative error could easily be relaxed, for example if there is knowledge that different measuring instruments were used for different objects.
\item 
    The assumed independence of errors across objects can also be relaxed, for example to accommodate spatial or temporal structure in the data, or hierarchical clusters.  The quasi-likelihood approach is flexible and is already well developed in this direction, especially following the influential work of \citet{liang+zeger1986}.
\end{enumerate}

\section*{Acknowledgment}
David Firth was supported by a Leverhulme Emeritus Fellowship, and Fiona Sammut by a PhD studentship from the U.K.~Engineering and Physical Sciences Research Council.

\bibliographystyle{apalike}
\bibliography{refs}

\end{document}